\documentstyle[twocolumn,prl,aps,epsfig]{revtex}

\begin{document}

\twocolumn[
\hsize\textwidth\columnwidth\hsize\csname@twocolumnfalse\endcsname
\draft
\title{Quasiparticles and c-axis coherent hopping in high $T_c$ superconductors }
\author{P. S. Cornaglia, K. Hallberg and C. A. Balseiro}
\address{Centro At\'omico Bariloche and Instituto Balseiro}
\address{Comisi\'on Nacional de Energ{\'\i}a At\'omica, 8400 S.C. de Bariloche, Argentina}
\date{Received \today }
\maketitle

\begin{abstract}
We study the problem of the low-energy quasiparticle spectrum of the extended {\it t-J} model and analyze the coherent hopping between weakly coupled planes described by this model. Starting with a two-band model describing the $CuO$ planes and the unoccupied
 bands associated to the metallic atoms located in between the planes, we obtain effective hopping matrix elements describing the {\it c}-axis charge transfer. A computational study of these processes shows an anomalously large charge anisotropy for doping 
concentrations around and below the optimal doping.  
\end{abstract}

]

\narrowtext

The anomalous charge transport of high $T_{c}$ materials is still among the
most important open problems related with the physics of these compounds\cite
{M-I,Dag-Rev}. It is now accepted that in order to build a complete theory
of high temperature superconductivity, a deep understanding of the normal
state properties of these compounds is needed and one of the most intriguing
questions concerns precisely the charge dynamics\cite{C-S}. The cuprate high 
$T_{c}$ superconductors have crystalline structures consisting of $CuO$
planes which makes all of them highly anisotropic materials. The electrical
conductivity reflects the anisotropy: while in the plane it shows a metallic 
behavior, along the {\it c}-axis in the underdoped regime, it indicates 
an incoherent charge transport;
moreover, it has an anomalous temperature behavior, and its frequency dependence
is not of the Drude type \cite{Homes}.

Band structure calculations together with a semiclassical theory of charge
transport, predict a metallic behavior in all directions with an anisotropy
much smaller than the experimental values. Electron-electron interactions
may strongly renormalize the bare parameters, however within the framework
of the Fermi-liquid theory the observed anisotropy cannot be understood. This led
Anderson and others to propose the failure of the conventional Fermi-liquid
theory in these compounds \cite{Anderson1}.

An important amount of experimental data has been accumulated during the
last years\cite{SI}. Among other things, the formation of charge stripes in
the normal state has been confirmed\cite{Stripes}. Whether this is relevant
for the occurrence of superconductivity is not yet clear, however it should
be taken into account in a detailed description of the normal phase.

Here we revisit the problem of the low-energy quasiparticle spectrum of the
extended {\it t-J} model and analyze the coherent hopping between weakly
coupled planes. Our starting point is a stack of $%
CuO$ planes coupled via the electron hopping $t_{\perp }({\bf k})$ to higher
energy intermediate states. The intermediate states represent the unoccupied
bands associated to the metallic atoms located in between the planes.

In a single-particle theory, the intermediate states can be easily
eliminated through a canonical transformation giving rise to an effective
hopping between consecutive $CuO$ planes of the form
$t_{\perp }^0({\bf k})=t_{\perp }^{2}({\bf k})/\Delta E_{{\bf k}}  \label{tmu} $, 
with $\Delta E_{\bf k}$ the one-particle energy difference between $%
{\bf k}$-states of the $CuO$ and the intermediate layer bands.
The principal symmetry of $t_{\perp }^0({\bf k})$ is obtained from
electronic structure calculations and for the case of $YBaCuO$ it is given
by $[\cos (k_{x}a)-\cos (k_{y}a)]^{2}$ \cite{LDA}.

In what follows we use an uncorrelated empty band to describe the
intermediate states and the extended {\it t-J} model to describe the $CuO$
planes. The extended {\it t-J} model including hole hopping to second\ and third
nearest-neighbor (NN) sites, and with {\it J/t} consistent with experimental
values, reproduces the photoemission results of the parent insulating
materials\cite{Kim,Dag2}. In the standard notation the model Hamiltonian
describing the plane $l$ is: 
\begin{equation}
H_{l}=J\sum_{\left\langle i,j\right\rangle }{\bf S}_{li}\cdot {\bf S}%
_{lj}-\sum_{n,m}t_{nm}c_{ln\sigma }^{\dagger }c_{lm\sigma }  \label{HtJ}
\end{equation}
where the matrix element $t_{nm}$ is $t$ for NN, $t^{\prime }$ for second NN
and $t^{\prime \prime }$ for third NN, $c_{lm\sigma }$ destroys an electron
with spin $\sigma $ at site $m$ and double occupation is not allowed. Using
the many-body states, we perform a canonical transformation that eliminates
the intermediate states and generates an effective Hamiltonian that includes
a charge transfer matrix element between two consecutive $CuO$ planes. Let
us first consider the case of a single hole. The two-plane states of the
system are indicated as $|\psi _{0,\nu }^{1}\rangle \otimes |\psi _{1,\nu
^{\prime }}^{2}\rangle $ where $|\psi _{m,\nu }^{l}\rangle $ is the 
many-body wave function of plane $l$ with $m$ holes and quantum numbers $\nu $
and the corresponding energy is $E_{m,\nu }$. We take the plane with no
holes in its ground state indicated by $\nu =0$. The effective matrix
elements connecting the states $|\psi _{0,0}^{1}\rangle \otimes |\psi _{1,\nu
}^{2}\rangle $ and $|\psi _{1,\nu }^{1}\rangle \otimes |\psi
_{0,0}^{2}\rangle $ accounts for a hopping - from one plane to the other -
of a single hole. The one-hole quantum numbers $\nu $ include the crystal
momentum ${\bf k}$, spin variables and other quantum numbers $\mu $ that
uniquely identify the state and in what follows we take $\nu \equiv ({\bf k}%
,\sigma ,\mu )$. Due to the momentum and spin conservation of the
Hamiltonian, the effective hopping of a hole with momentum ${\bf k}$ and
spin $\sigma $ is given by:

\begin{equation}
\widetilde{t}_{\perp }({\bf k})=\frac{t_{\perp }^{2}({\bf k})M_{1}({\bf k}%
,\sigma ,\mu )M_{2}^{*}({\bf k},\sigma ,\mu )}{E_{0,0}-E_{1,({\bf k},\sigma
,\mu )}-\varepsilon _{{\bf k}\sigma }}  \label{teff}
\end{equation}
where $\varepsilon _{{\bf k}\sigma }$ is the energy to add an electron at
the intermediate band, $M_{1}({\bf k},\sigma ,\mu )=\langle \psi _{1,({\bf k}%
,\sigma ,\mu )}^{1}|c_{1{\bf k}\sigma }|\psi _{0,0}^{1}\rangle $ and $%
M_{2}^{*}({\bf k},\sigma ,\mu )=\langle \psi _{0,0}^{2}|c_{2{\bf k}\sigma
}^{\dagger }|\psi _{1,({\bf k},\sigma ,\mu )}^{2}\rangle $ are the matrix
element for the creation and destruction of a hole in the planes. We
calculate many-body energies and matrix elements by exact diagonalization of
 small clusters with periodic boundary conditions.

\begin{figure}[t]
\epsfysize = 5.5cm
\begin{center}
\leavevmode
\epsffile{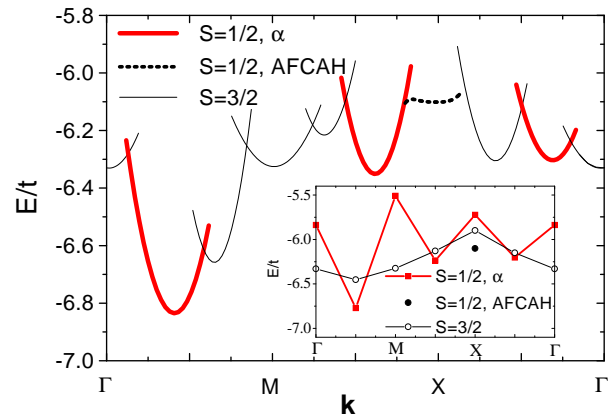}
\end{center}
\caption{Dispersion of one hole for the extended {\it t-J} model on a $4\times 4$ cluster; $t^\prime/t=-0.35$, $t^{\prime\prime}/t=0.25$, and $J/t=0.4$. In the inset the one hole PES is shown together with the $S=3/2$ and the AFCAH state energies. Here $\Gamma\equiv (0,0)$, 
$M \equiv (\pi,\pi)$, and $X\equiv (\pi,0)$.}
\label{fig1}
\end{figure}

Consider the one-hole ground state in a 16-site cluster: this state has momentum $%
{\bf k}=(\pi /2,\pi /2)$ and the product of the matrix elements in Eq.(\ref
{teff}) is the weight of the quasiparticle peak ${\it Z}_{k}$. Since the
same factor ${\it Z}_{k}$ renormalizes the in-plane hole mass, for one hole
close to the Fermi surface, interactions do not renormalize the anisotropy.
The one-hole state with ${\bf k}=(\pi /2,\pi /2)$ can be viewed as a spin-$%
1/2$ and charge $e$ quasiparticle, where the spins across the hole have
ferromagnetic correlations. To distinguish this excitation from others that
appear at different regions of the Brillouin zone (BZ), we refer to it as
the $\alpha $-type spin-$1/2$ quasiparticle. This quasiparticle exists for
all the momenta in the BZ and its energy can be obtained simply by following
the lowest energy peak of the one-particle spectral density,{\it \ i.e.} its
dispersion relation, shown in Fig.(\ref{fig1}), gives the one-hole photoemission
(PES) dispersion of the insulator. It is important to note that this
quasiparticle is not the lowest energy state for all momenta ${\bf k}$. For
realistic parameters that fit the PES ($J/t=0.4$ , $t^{\prime }/t=-0.35$ and 
$t^{\prime \prime }/t=0.25$), the lowest energy states in the $4\times 4$
cluster with ${\bf k}=(0,0)$ and $(\pi ,\pi )$ have total spin $3/2$\cite
{Aligia}. These high spin states have been obtained in clusters of different
sizes (up to 20 sites) and for different parameters. However it is not clear
that in the thermodynamic limit they will remain being the lowest energy
excitations around the $\Gamma $ and $M$ points. These states can be
interpreted as high spin quasiparticles that occur as a precursor of the
Nagaoka state. The magnetization is confined around the hole like in a
polarized spin bag. These states are not connected to the ground state of
the undoped system when a hole is created; in other words the matrix
elements in Eq.(\ref{teff}) are zero and these excitations do not propagate
- to lowest order - from one plane to the other. The reason is that when a
hole is created on the spin-zero ground state of the insulating system, the
resulting states have total spin $1/2$, and the spin wave function is
orthogonal to the spin-$3/2$ excitations.

\begin{figure}[t]
\epsfysize = 5.5cm
\begin{center}
\leavevmode
\epsffile{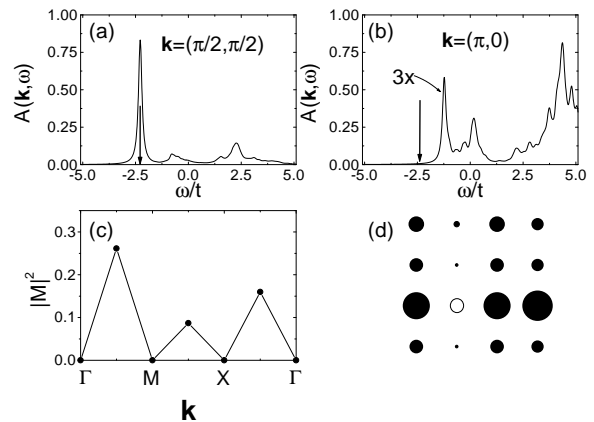}
\end{center}
\caption{(a),(b) Single hole spectral functions $A({\bf k}, \omega)$, for the extended t-J model, same parameters as Fig.(1) (broadening $\delta=0.1t$). The arrow points to the one-hole lowest energy state with the corresponding momentum. (c) Matrix element $|M|^2$ for the available momenta on a $4\times 4$ cluster. (d) Hole density around one hole (open circle) on the two holes ground state (${\bf k}=(0,\pi)$). Full circles radius are proportional to the hole density. }
\label{fig2}
\end{figure}

More interesting are the one-hole lowest energy states at ${\bf k}=(\pi ,0)$
or $(0,\pi )$. These states have been studied recently in some detail\cite
{Dag2,Jap}. The spin correlations across the hole are antiferromagnetic
(AFCAH), like in 1D systems, and it has been speculated that this is an
indication of charge and spin separation in the 2D {\it t-J} model\cite{Dag3}.
Similar to the high spin excitation case, here also, the matrix elements
connecting this state with the ground state of the undoped system by
creating a hole is zero. Although the total spin of the ${\bf k}=(\pi ,0)$
lowest energy excitation is $1/2$, the symmetry of its spin wave function is
orthogonal to that of $c_{1{\bf k}\sigma }|\psi _{0,0}^{1}\rangle $. The
one-hole spectral density of the insulator does not show any structure at
the energy of this state as shown in Fig. (\ref{fig2}b). Again, to lowest order, this 
excitation does not propagate from one plane to the other. This is a property 
of the $tt^\prime t^{\prime \prime}-J$ model and it is not observed in the 
{\it t-J} model for similar values of {\it J/t}.
\\
In order to
qualitatively determine the regions of the BZ where the different type of
excitations are the lowest energy ones, we have continuously changed the
boundary conditions in the cluster and the results are shown in Fig. (1). In
a region around the $X$ point, the spin-$1/2$ states with AFCAH are stable
(dashed line). If these excitations actually describe a region of charge and
spin separation, the dashed line branch does not correspond to the
dispersion of a single quasiparticle but to the superposition of a holon and
a spinon. It has been suggested that these states play a central role in the
many-holes states and may be responsible for the formation of stripes\cite
{Dag3}. As we show below, they are also relevant for the coherent hopping
along the {\it c}-axis in the doped systems.

Summarizing the one-hole results, we have shown that in different regions of
the BZ, the lowest energy excitations are of different nature. Close to the
Fermi surface (${\bf k}=(\pi /2,\pi /2)$), there is a spin-$1/2$
quasiparticle that propagates along the {\it c}-axis with an effective hopping
renormalized with the ${\it Z}_{k_{F}}$ factor. At the $X$ point, the anomalous
spin-$1/2$ excitation with AFCAH is the lowest energy state. A hole in this
state is confined, up to second order in $t_{\perp }$, in a single
plane. Here the word confinement is used in the sense that there is no
bonding- antibonding splitting of the degenerate two-plane lowest energy
state. In Fig.(\ref{fig2}c) the matrix element for different values of ${\bf k}$ is
shown. Higher order terms may generate a coherent hopping for this type of
excitations. In any case, we expect these higher order
terms to generate a very small dispersion along the {\it c}-axis \cite{Com2}. 
Moreover, it should be noted that for the 20-site cluster, for 
the ${\bf k}=(0,\pi)$ state 
with robust AFCAH, the second order matrix elements non zero but very small, 
{\it i.e.} the quasiparticle peak in the spectral density of the insulator 
has an extremely small weight \cite{Dag2,Jap}, leading to a strongly
renormalized {\it c}-axis hopping for this state, in qualitative 
agreement with the results for the 16-site system. The non-crossing approximation 
in larger systems also gives zero weight \cite{Aligia}.

Let us now turn to the many-holes states. For the same parameters as
above, the two-hole ground state of the $4\times 4$ cluster has
total momentum ${\bf k}=(0,\pi )$ or $(\pi ,0)$ and spin zero. In these states, the
hole-hole correlation function $\langle n_{i}n_{j}\rangle $ with $n_{i}$ the
hole number at site $i$, clearly shows the tendency to stripe formation. For
the state with ${\bf k}=(0,\pi )$ [$(\pi ,0)$], the holes are mostly aligned
in the same row along the $x$ [$y$] direction as shown in Fig (2d). These
results are in agreement with previous results obtained in ladders and
clusters. In the 18-site cluster, stripes are observed in the three 
holes ground state \cite{Dag3}.
Based on the spin correlations across the hole, it has been argued
that these stripes are made out of one-hole building-blocks with AFCAH\cite
{Dag3}. In other words, the two holes wave function has a large component of
the one-hole wave function of the ${\bf k}=(0,\pi )$ type. This could
strongly influence the charge mobility along the {\it c}-axis. We first analyze
the case of two $4 \times 4$ planes with three holes corresponding to a doping $x\simeq
0.094$. With a straightforward extension of the previous treatment and
notation, we calculate the effective matrix element mixing the two
(degenerate) states with wavefunctions $|\psi _{1,\nu }^{1}\rangle \otimes
|\psi _{2,\xi }^{2}\rangle $ and $|\psi _{2,\xi }^{1}\rangle \otimes |\psi
_{1,\nu }^{2}\rangle $ which now are $M_{1}({\bf k},\nu ,\xi )=\langle \psi
_{2,\xi }^{1}|c_{1{\bf k}\sigma }|\psi _{1,\nu }^{1}\rangle $ and $M_{2}^{*}(%
{\bf k},\nu ,\xi )=\langle \psi _{1,\nu }^{2}|c_{2{\bf k}\sigma }^{\dagger
}|\psi _{2,\xi }^{2}\rangle $. For these matrix elements to be non zero, the
operators $c_{1{\bf k}\sigma }$ and $c_{2{\bf k}\sigma }^{\dagger }$ have to
be chosen to conserve total momentum and spin. 

If the two planes are in their ground state, the product of
these matrix elements is smaller than for the one-hole case. 
This result, is consistent with the conjecture that the two holes wave function is
mainly made out of one-hole building blocks with AFCAH since, as we showed
above, this type of excitation tends to be confined in one plane. This effect is largely 
enhanced when doping increases as in the case of five holes in two $4\times 4$ planes ($%
x\simeq 0.156$). In this case, if the planes are in the ground state, with
two and three holes respectively, the effective matrix element to transfer a
charge from one plane to the other is exactly zero and to lowest order,
charge is confined in the planes. In fact as
shown in Fig. (3b), the PES of the system with two holes has no structure at
the energy of the three-hole ground state. The case of four holes ($x=0.125$%
) deserves a special comment: the ground state of the uncoupled planes has
two holes in each one and the hopping mixes it with excited states. It has been proposed to artificially shift the energy levels
of the states with one hole in one plane and three in the other to recover a
degeneracy\cite{8x8}. These shifts are unimportant when calculating the
matrix elements which are $\langle \psi _{3,\xi _{0}}^{1}|c_{1{\bf k}\sigma
}|\psi _{2,\nu _{0}}^{1}\rangle \times \langle \psi _{1,\eta _{0}}^{2}|c_{2{\bf k%
}\sigma }^{\dagger }|\psi _{2,\nu _{0}}^{2}\rangle $ with the indices $\xi
_{0}$, $\eta _{0}$ and $\nu _{0}$ corresponding to the ground state of a
plane with the indicated number of holes. These matrix elements are
precisely the ones calculated in the two previous cases and the product is
zero. For the case of 6 holes in two $4\times 4$ planes we reach
similar conclusions.
\\

Finally we have evaluated the effective hopping for large number of holes, $9
$ and $11$ holes in the two planes which correspond to dopings $x$ larger
than the optimal doping. In this region, the effective hopping calculated
always in second order in $t_{\bot }$ and in the ground state, are non-zero.

We have also studied the effective hopping along the {\it c}-axis for 18-site and 20-site clusters, 
and small doping. For the 18-site cluster, and the same parameters as before, the 
two holes low energy spectrum presents a quasidegeneracy between the spin zero and spin one states. The effective hopping calculated in the lowest energy $S=0$ state, which we believe is the relevant in the thermodynamic limit, shows the same behavior as in the $4\times 4$ cluster, indicating that our results are not dominated by the particular symmetry of these clusters.

\begin{figure}[t]
\epsfysize = 5.5cm
\begin{center}
\leavevmode
\epsffile{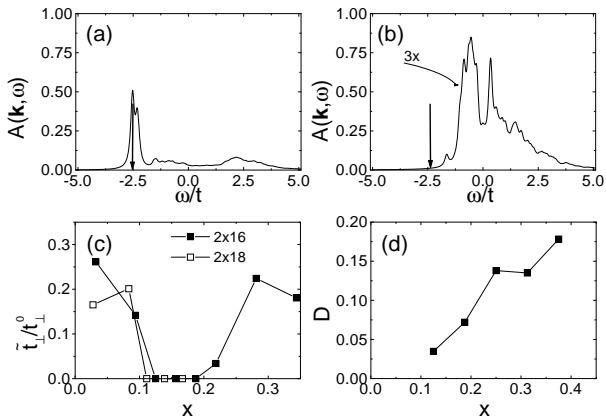}
\end{center}
\caption{(a) Single-hole spectral function $A({\bf k}, \omega)$ for a $4 \times 4$ cluster with one hole (${\bf k}=(-\pi/2, \pi/2)$), for the extended t-J model and  same parameters as Fig (1). The arrow points 
 to the final state energy. (b) Same as (a) for a system with two holes (${\bf k}=(0,0)$).  (c) Effective hopping between $CuO$ planes up to second order in $t_\perp$ as a function of the hole doping $x$ for the 16-site and 18-site clusters. (d) Drude weight averaged over both directions for the $4 \times 4$ cluster as a function of the hole doping $x$.}
\label{fig3}
\end{figure}

In Fig. \ref{fig3}(c) we summarize the results for the matrix elements, which give an
estimate of ratio $\widetilde{t}_{\bot }/t_{\bot}^0$ as a
function of the doping $x$, here with $t_{\bot}^0$ is
the bare hopping obtained from band structure calculations.
Our results show that while a single hole propagates
along the {\it c}-axis, for doping close and smaller than the optimal doping $%
x_{op}\simeq 0.2$ the lowest order hopping evaluated in the ground state
cancels\cite{Rojo} and increases for $x>x_{op}$. This behaviour suggest that when stripes are stabilized,
the {\it c}-axis hopping reduces to very small values. These results are in agreement with 
previous calculations \cite{Toh2} which suggest that the weight of the quasiparticle 
peak decreases in the presence of stripes. As we mentioned
above, higher order terms generate a non-zero hopping. However, at least in
the one-hole case in the two $\sqrt{8}\times \sqrt{8}$ planes\cite{Com2}, these
contributions are very small reducing the bare hopping by a factor smaller
than $10^{-3}$. 
\\

For the sake of completeness we have also calculated the in-plane Drude 
weight $D$ for the same parameters as before. As shown in Fig. \ref{fig3}(d) $
D$ increases with doping (as in the t-J model \cite{Dag-Rev}) in the region 
were $\widetilde{t}_{\bot }$ presents a marked depletion.  
Our treatment, based in a canonical transformation and the estimation of
matrix elements by exact diagonalization of small clusters is not
appropriate to give a definite answer to what may be the most relevant
question related to the charge dynamics in high $T_{c}$ materials, that is
whether under some conditions real confinement of charge excitations in the $%
CuO$ planes is obtained. However our results are consistent with, and can
explain, the available experimental data. The behavior here obtained is 
robust under changes of the parameters within the accepted range of values.
  On one hand, to lowest order in $%
t_{\bot }$, in the ground state charge is confined for doping $x$ close to
and below the optimal doping $x_{op}$. Higher order terms in $t_{\bot }$ may
generate a small coherent hopping $\widetilde{t}_{\bot }\simeq t_{\bot}\times 
10^{-3}$. Estimations of the bare hopping using LDA band
structure calculations give $t_{\bot}=0.25$ eV for the case of bilayers
and $t_{\bot}\simeq 0.03$ eV for the interplane matrix element\cite
{LDA}. That means that down to very low temperatures ($kT \sim \widetilde{t}_{\bot}$),  we expect a diffusive-like propagation along the {\it c}-axis with an
anomalous temperature dependence. On the other hand, the small cluster
calculations are remarkable in the sense that they predict an anomalously
large charge anisotropy precisely in the region where superconductivity is
stable. This agrees with the experimental data\cite{tamasaku} and
strengthens the ideas developed by Anderson and coworkers \cite{and-science}
that argued that the lack of single particle intraplane coherent hopping
enhances superconductivity.

We thank A.A. Aligia, A. Ceccatto, and B. Normand for helpfull discussions. This work was parcially supported by the CONICET, ANPCYT(02151).


\begin{references}
\bibitem{Com2} {We have checked this in a system consisting of two $\sqrt{8}\times \sqrt{8}$ coupled planes, which has the three types of excitations shown in Fig. (1). The splitting between the bonding and antibonding states is of the order of $Z_k t_\perp$ for the $\alpha$-type, $S=1/2$ quasiparticle, while it is three orders of magnitude smaller for the other two excitations.}
\end{references}
\end{document}